\documentclass[reprint,twocolumn,aps,prl,superscriptaddress,showpacs,amsmath,amssymb]{revtex4}

\usepackage{graphicx,amssymb,amsmath,mathrsfs,mathtools,siunitx}
\usepackage{tabularx,verbatim}
\usepackage[normalem]{ulem}
\usepackage{color}

\newcommand{\Sr}{{Sr$_2$RuO$_4$}}
\newcommand{\HQV}{{half-flux-quantum}}
\newcommand{\FQV}{{full-flux-quantum}}
\newcommand{\TME}{{torque magnetometry experiment}}

\begin{document}

\bibliographystyle{apsrev}

\title{Magnetoresistance oscillations and the half-flux-quantum state in spin-triplet superconductor Sr$_2$RuO$_4$}

\author{X. Cai}

\author{Y. A. Ying}

\affiliation{Department of Physics and Materials Research Institute, Pennsylvania State University, University Park, PA 16802, USA}

\author{J. E. Ortmann}
\author{W.-F. Sun}
\author{Z.-Q. Mao}

\affiliation{Department of Physics, Tulane University, New Orleans, LA 70118, USA}

\author{Y. Liu}
\email{yxl15@psu.edu}

\affiliation{Department of Physics and Materials Research Institute, Pennsylvania State University, University Park, PA 16802, USA}
\affiliation{Department of Physics and Astronomy, Shanghai Jiao Tong University, 800 Dong Chuan Road, Shanghai 200240, China}
\affiliation{Collaborative Innovation Center of Advanced Microstructures, Nanjing 210093, China}

\date{\today}

\begin{abstract}

We report results of our low-temperature magneto electric transport measurements on micron-sized short cylinders of 
odd-parity, spin-triplet superconductor Sr$_2$RuO$_4$ with the cylinder axis along the $c$ axis. The in-plane magnetic field and 
measurement current dependent magnetoresistance oscillations were found to feature an amplitude much larger than that 
expected from the conventional Little-Parks effect, suggesting that the magnetoresistance oscillations originate from vortex crossing. The free-energy barrier that controls the vortex crossing was modulated by the magnetic flux enclosed in the cylinder, the in-plane field, measurement current, and structural factors. Distinct features on magnetoresistance peaks were found, which we argue to be related to the emergence of half-flux quantum states, but only in samples for which the vortex crossing is confined at specific parts of the sample. 

\end{abstract}

\maketitle

%%%%%%%%%%%%%%%%%%%%%%%%%%%%%%%%%%%%%%%%%%%%%%%%%%%%%%%%%%%%%%%%%%%%%%%%

Fluxoid quantization in a doubly connected conventional superconductor in the unit of a full flux-quantum, $\Phi_0=hc$/2$e$ (where $h$ is the Plank constant and $e$ the electron charge), is a direct consequence of pairing of the electrons and the emergence of long-range phase coherence among the paired electrons.\cite{Tinkham}
Deaver and Fairbank\cite{DeaverFluxQuantization} and  Doll and N$\ddot{\textrm{a}}$bauer\cite{DollFluxQuantization} measured the magnetic flux trapped in a superconducting hollow cylinder and  the torque on the circulating supercurrent, respectively, to demonstrate this remarkable effect.
The physics of this effect was further clarified by the Little and Parks (LP) experiment\cite{LittleParks1962}, demonstrating the oscillations in the superconducting transition temperature, $T_c$, and the periodic variation of the free energy with the applied magnetic flux in a doubly connected superconducting cylinder. 
Therefore, after the initial experimental evidence for the half-flux-quantum vortices\cite{Kee2000PRB} was found in the odd-parity, spin-triplet superconductor \Sr\cite{RiceSigristJPCM1995,BaskaranPhysicaB1996,MackenzieMaenoRMP2003,MaenoJPSJ2012,LiuMaoPhysicaC2015}  in the \TME\cite{Jang2011Science}, 
the LP measurement has been highly desirable so as to obtain insights into the physics of the half-flux-quantum state in this unconventional superconductor.

The \HQV\ state is allowed in a spin-triplet superconductor along with the conventional full-flux-quantum state because of the presence of spin and orbital parts of the order parameter. A possible scenario is that a phase winding of $\pi$ is formed separately in each part of the order parameter around a doubly connected sample or a vortex core to maintain the singlevaluedness of the order parameter. Only the circulating supercurrent due to the orbital phase winding features a magnetic flux, giving rise to a vortex state featuring a \HQV\ of $\Phi_0$/2. The free energy of the half-flux-quantum state is usually higher than the conventional full-flux quantum one, making its physical realization difficult. For a doubly connected, micron-sized crystal of Sr$_2$RuO$_4$, the free energy of the \HQV\ state appeared to be lowered near applied half-flux quanta by the application of an in-plane magnetic field\cite{Jang2011Science} as proposed theoretically\cite{VakaryukLeggett2009PRL}.

Recently we carried out magnetoresistance oscillation measurements on micron-sized, single crystal rings of Sr$_2$RuO$_4$\cite{Cai2013PRB}, in which a large number of pronounced resistance oscillations with an amplitude much larger than that expected from the LP effect were observed down to very low temperatures. 
The observed magnetoresistance oscillations were attributed to $c$-axis vortices moving across the sample that leads to a finite transverse voltage according to the Josephson relation\cite{Tinkham}.
The oscillation amplitude is controlled by the barrier potential for vortex crossing, which is a function of applied flux due to fluxoid quantization.  
No features associated with \HQV\ states were found over a wide range of the out-of-plane field without the application of an in-plane field.
Interestingly, a fit of our data 
%[Supplementary information] 
to the Ambegaokar-Halperin (AH) model of thermally activated vortex crossing over a free energy barrier\cite{SochnikovNNano,VakaryukVinokur2011} yielded values of zero-temperature penetration depth much larger than that of the bulk\cite{RobertsDissertation}, which is consistent with the small magnetic moments observed in the torque magnetometry experiment compared 
to the numerical calculation\cite{Roberts2013PRB}. Here we present magnetoresistance oscillation measurements in the presence of 
an in-plane field with varying measurement currents and magnetoresistance signatures of the half-flux-quantum state, thereby providing 
insights into conditions favoring experimental observation of this exotic state.  

%%%%%%%%%%%%%%%%%%%%%%%%%%%%%%%%%%%%%%%%%%%%%%%%%%%%%%%%%%

To prepare our samples, thin crystal plates of \Sr\ were made on a Si/SiO$_2$ substrate by mechanical exfoliation from a bulk single crystal. 
Four- or six-point electrical leads, made of 200 nm Au and an underlay of 10 nm Ti, were prepared on the crystal plates by photolithography. A doubly connected cylinder with four 
leads was cut from the crystal plates using a focused ion beam (FIB), with Sr$_2$RuO$_4$ leads extending from the cylinder to the Ti/Au contacts.
Fig.~1a shows a scanning electron microscopy (SEM) image of a typical cylinder, which has a height of about 0.74 $\mu$m and a wall thickness increasing from top to bottom 
due to the profile of the ion beam. This hollow cylinder has a mean wall-thickness $(w)$ of 0.26 $\mu$m and a mean radius $(r)$ of 0.58 $\mu$m as indicated in Fig.~1b.
More details in sample fabrication and the estimation of sample 
dimensions can be found in reference~\cite{Cai2013PRB}. Our samples were measured in a dilution refrigerator with a base temperature 
of 20 mK, using a dc technique. Since both the out-of- and the in-plane fields, $H_{||c}$ and $H_{||ab}$, are needed, a homemade 
superconducting Helmholtz coil was incorporated inside a large superconducting solenoid.

Almost all Sr$_2$RuO$_4$ samples prepared this way were found to be superconducting with an onset $T_c$ sample dependent. 
In Fig.~1c, the temperature dependent resistance data
 revealed a broad transition with a zero-resistance 
$T_c$ dependent on the measurement current. A resistance peak was found around 1.6~K, an anomaly observed previously 
in superconducting nanowires and attributed to charge imbalance encountered when superconducting voltage leads were 
used.~\cite{SanthanamPRL1991} The onset $T_c$ of these samples is higher than the bulk phase, 1.5 K, likely due to the presence 
of dislocations in the thin crystal plates of \Sr.~\cite{Ying2013Ncomms}

\begin{figure}[t!]
\centering
\includegraphics[natwidth=243,natheight=216]{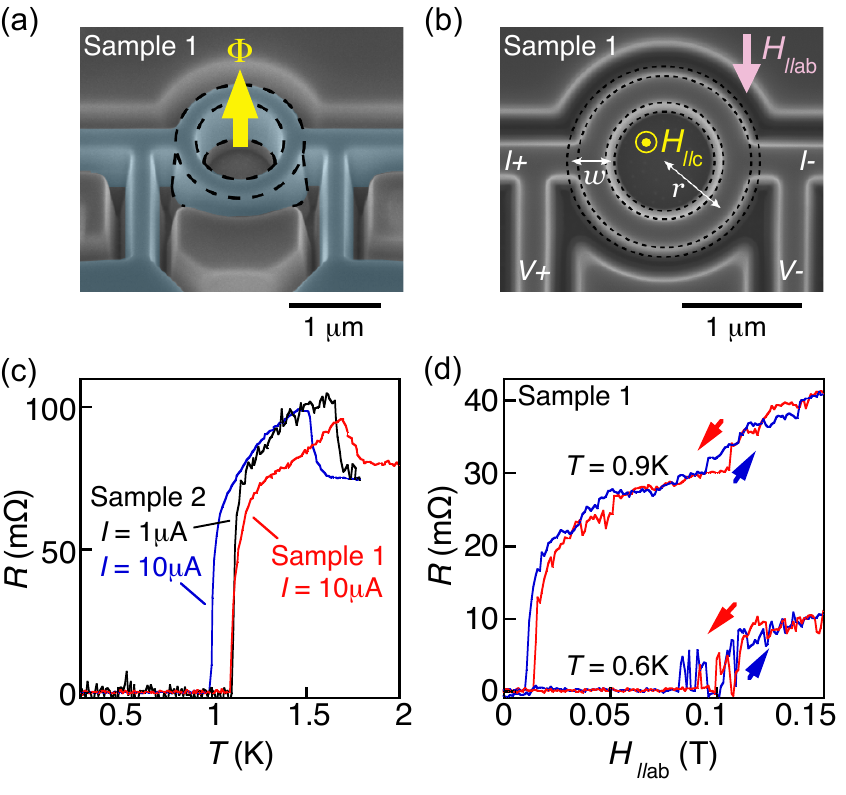}
\caption{(a) A false-color SEM image of Sample 1 prepared on a Si/SiO$_2$ substrate. The cylinder and the leads (in blue) were covered by 
a layer of SiO$_2$ used to protect the crystal during the FIB cutting. The image was taken at a \SI{30}{\degree} angle from the axial 
direction. The slightly increased wall thickness towards the bottom of the cylinder is evident. 
(b) A SEM image of the sample in (a) showing the measurement setup. The outer edges of the cylinder on the top 
and the bottom surfaces of the crystal plate are indicated. 
(c) Temperature dependence of sample resistance, $R$ vs. $T$, measured for two samples with applied currents, $I$, as indicated. 
Sample 2 has the same $w$ and $r$ as Sample 1 and a height of about 0.78 $\mu$m.
(d) Sample resistance as the function of the in-plane field, $R$ vs. $H_{||ab}$, at two different temperatures with $I$ = 30$\mu$A. 
Arrows indicate the field sweeping directions.
}
\label{opts}
\end{figure}

We measured the sample resistance as a function of the in-plane field, $H_{||ab}$, aligned perpendicular to the current leads (Fig.~1b). 
Hysteresis as well as fluctuations were observed (Fig.~1d), especially for the curves measured at a low temperature, 0.6 K, possibly due 
to the depinning of parallel vortices.
% [Supplementary information]. 
Even though the in-plane field could be misaligned from the $ab$-plane of the cyliners by a small angle, 
the vortices in the sample should still be aligned in the in-plane direction because of the strong anisotropy of Sr$_2$RuO$_4$. 
%The motion of depinned parallel vortices will lead to a finite voltage according to the Josephson relation and therefore finite sample 
%resistance, resulting in observed fluctuation and hysteresis.
The critical field, at which the sample resistance becomes non-zero, was found to decrease with the increasing measurement current. 
It should be noted that the $c$-axis vortices can appear simultaneously in \Sr\ crystals.\cite{Dolocan2006PRB,Dolocan2005PRL} 
%The motion of all vortices across the sample causes finite transverse voltage and therefore sample resistance.

%%%%%%%%%%%%%%%%%%%%%%%%%%%%%%%%%%%%%%%%%%%%%%%%%%

\begin{figure}[t!]
\centering
\includegraphics[natwidth=243,natheight=216]{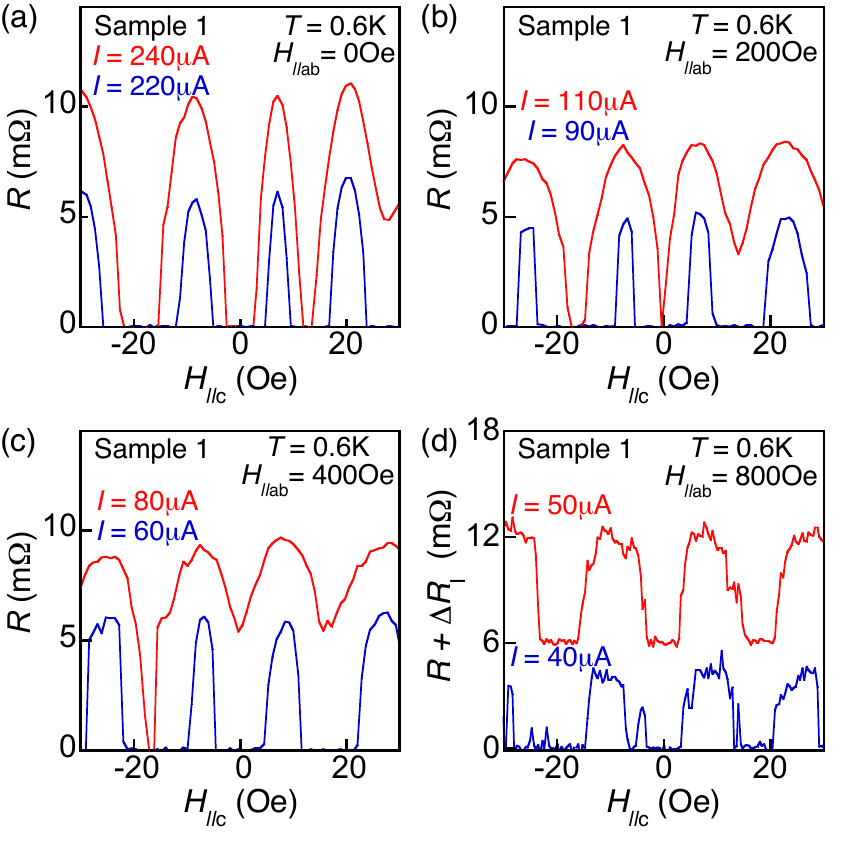}
\caption{
Magnetoresistance oscillations, $R$ vs. $H_{||c}$, for Sample 1 measured at $T$=0.6 K far below zero-field $T_c$ for an in-plane field 
of 0 (a), 200 (b), 400 (c) and 800 G (d), respectively. The measurement currents are as indicated. The top curve in (d) is shifted vertically by 6 m$\Omega$ for clarity.}
% (e) and (f) $R$ vs. $H_{||c}$, for Sample 2 measured under different conditions as indicated. The top curve in (f) shifted by 20 m$\Omega$ for clarity.  }
\label{opts}
\end{figure}

Pronounced magnetoresistance oscillations were observed at fixed temperatures as the out-of-plane field, $H_{||c}$, was ramped up (Fig.~2). The oscillation period is $\Phi_0$ based on the sample dimensions. 
A large measurement current was used in order for the magnetoresistance oscillations to be measured at such low temperatures.
The vortex crossing origin of the magnetoresistance oscillations is confirmed by a quantitative comparison between 
the experimental values of the oscillation amplitude and that expected from the conventional LP effect.
No obvious feature corresponding to the transition between half- and full-flux quantum states in the resistance oscillations 
was seen in the presence of a constant in-plane magnetic field, $H_{||ab}$, up to 800 Oe. 
At such a high in-plane field, the magnetoresistance oscillations showed significant irregularities and instability (Fig.~2d).

%%%%%%%%%%%%%%%%%%%%%%%%%%%%%%%%%%%%%%%%%%%%%%

 \begin{figure}[t!]
\centering
\includegraphics[natwidth=243,natheight=216]{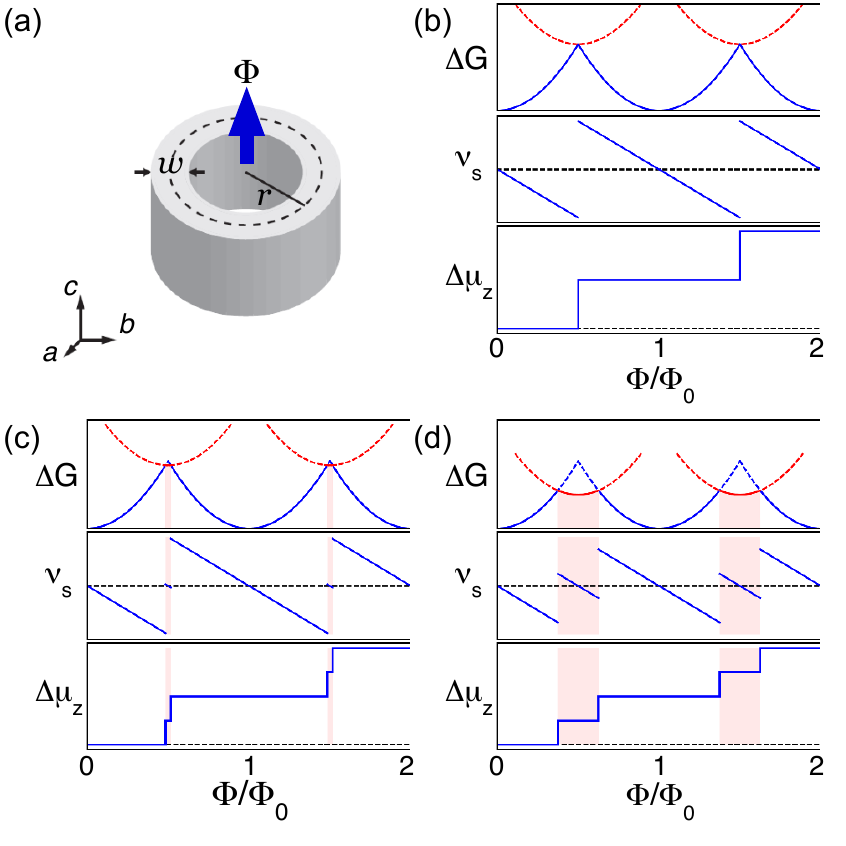}
\caption{
%\textcolor{blue}{
Superfluid velocity $v_s$, variation of magnetization $\Delta \mu_z$, and the kinetic part of the Gibbs free energy, $\Delta$G,  
as a function of $\Phi$ calculated from equation (1) for a doubly connected cylinder with $r=0.58$ $\mu$m and $w=0.26$ $\mu$m, 
as shown in (a), at a temperature close to $T_c$ (b) and 0.6 K (c); Near $T_c$, we assume $\rho_{sp}/\rho_s\rightarrow1$ 
and $\lambda\rightarrow\infty$; Values of $\rho_{sp}/\rho_s$ =0.25 and $\lambda\sim$166 nm are used for $T$=0.6~K. 
Values of $\Delta$G for the full- and half-flux-quantum states are plotted in blue and red, respectively. 
The light red shadings present the stability regions of the \HQV\ state; 
(d) When the width of the stability region is $\Phi_0$/4, the energy difference between the full- and the half-flux-quantum states 
at $\Phi=\Phi_0 /2$ is 0.31, and the depth of the free energy dip is only 0.04. 
The scale for $\Delta$G is 0.4($\Phi_0^2/8\pi^2r^2$)$\beta/(1+\beta)$.
 }
\label{opts}
\end{figure}

To understand the data presented above, the magnetoresistance oscillation signature of the \HQV\ states in our samples 
needs to be analyzed. for simplicity, our sample is modeled as a thin-wall hollow cylinder without leads  (Fig.~3a).
Starting with a two-component order parameter, the Gibbs free energy 
per unit length of a cylinder with a wall thickness of $w$ and a radius of $r$ in the thin-wall limit ($w\ll r, \lambda$) 
as a function of an external field, $H_{||c}$, has the following form,
\cite{Chung2007PRL,VakaryukLeggett2009PRL}
\begin{equation}
\begin{split}
& G(n_s, n_{sp})=\left(\frac{\Phi_0^2}{8\pi^2 r^2}\right) \\
& \left(\frac{\beta}{1+\beta}\left[ \left(n_s-\frac{\Phi}{\Phi_0}\right)^2+n^2_{sp}(1+\beta)\frac{\rho_{sp}}{\rho_s} \right]
-\left(\frac{\Phi}{\Phi_0}\right)^2\right) 
%+ g_{so}
\end{split}
\end{equation}
where $\beta=rw/2\lambda^2$ is the screening parameter, $\Phi=\pi r^2 H_{||c}$ is the applied flux through the cylinder, $n_{s}$ and $n_{sp}$ are the winding numbers for orbital and spin parts of the order parameter, $\rho_{s}$ and $\rho_{sp}$ are charge and spin supercurrent densities.
A \HQV\ state is in general energetically less favorable than a \FQV\ state due to the presence of unscreened spin supercurrent, indicated by the $n_{sp}$ term in equation (1).
The stability condition for \HQV\ states is given as $(1+\beta)({\rho_{sp}}/{\rho_s})<1$, which makes
the total kinetic energy of the spin and charge currents in the half-flux-quantum state, $\Delta G(l_s=\pm1/2,l_{sp}=1/2)$, lower than that of the full-flux-quantum state in the vicinity of $\Phi=\Phi_0/2$, $\Delta G(0,0)$ or $\Delta G(\pm 1,0)$. 
The sample size needs to be sufficiently small, that means a small $\beta$, for a given $\rho_{sp}/\rho_s$ in order to reduce the energy cost of the spin supercurrent. 
The magnetization curves, $\Delta \mu_z$, can be obtained by taking the derivative of the free energy with respect to the applied magnetic field, with the linear background subtracted.
It should be noted that the above analysis includes the contribution from neither the spin-orbit coupling, $g_{so}$ nor the superconducting electrical leads. The latter contributes to the Gibbs free energy through the absence of the screening of the spin supercurrent and the spin-orbital coupling. 

The value of $\rho_{sp}/\rho_s$ is not known for Sr$_2$RuO$_4$. Numerical studies on the stability regions of half-flux-quantum states 
based on the torque magnetometry data suggested a value of $\rho_{sp}/\rho_s=0.25$ at 0.6 K.\cite{Roberts2013PRB} 
%Based on our sample dimensions, the stability region of half-flux-quantum states is narrow without the application of an in-plane field (red shading in Fig.~3c).
The ground state free energy shown in Fig.~3 is calculated using parameters of bulk \Sr. 
Using the AH model to fit our data we obtain a large penetration depth, and therefore a $\beta$ value much smaller than expected 
from that using parameters of the bulk, which favors the presence of the \HQV\ state; however, a large value of the penetration depth 
also suggests a suppression of superconductivity which indicates that
$\rho_{sp}/\rho_s$ is likely to be much lager than that used in the numerical study, close to 1. The free energy contribution from in-plane 
field is given as $\Delta F=-g\mu_B (\rho_{\uparrow}-\rho_{\downarrow})|B_{||ab}|$.\cite{Roberts2013PRB} Assuming that 
$\rho_{\uparrow, \downarrow}$ is independent of $\Phi$, the free energy lowering of the half-flux-quantum state is proportional to the 
magnitude of the in-plane field. Fig.~3d is obtained by directly shifting the free-energy curve of \HQV\ state vertically to match the width 
of the stability region observed in the torque magnetometry measurements\cite{Jang2011Science}. 
The free energy variation in the half-flux-quantum state is tiny compared to the free energy difference between the full- and half-flux-quantum states. 
The sample resistance tracks the free-energy monotonically even though the precise form that is associated with vortex crossing
%, which is associated with vortex crossing, 
is not known.
% for a spin-triplet superconductor. 

With the emergence of the \HQV\ state we expect the resistance peaks of $\Phi_0$ oscillations to develop a dip around applied half-flux quanta,
%$\Phi=\Phi_0/2$, 
especially as the stability regime becomes substantial.  
%We expect that the resistance peaks to feature a flat top even though the two corners marking the entrance/exit of the half-flux-quantum state may not translate into corresponding features in the resistance peak. 
%Interestingly, in high in-plane fields, the resistance peaks shown in Fig.~2d appear to become flatter than those seen in the smaller in-plane fields (Fig.~2a-c). However, this is not a sharp enough feature to be tied uniquely to the existence of the half-flux-quantum state.
%In addition, f
For a sample used in our experiment, vortex trapping and pinning within the sample, the free energy contribution from the electrical leads, and the spatial variation of the superconducting order parameter will all complicate the free-energy barrier for the vortex crossing, making sample resistance as a function of the magnetic flux difficult to track.  In the torque magnetometry experiment, the entry of parallel vortices was found to be accompanied by the field shift of the transitions between the adjacent fluxoid states and a reduction in the stability region of the half-flux quantum states. In our experiment, the resistance peaks seen in high in-plane fields (Fig.~2d) appear to be flatter than those seen in the smaller in-plane fields (Fig.~2a-c), consistent with theoretical expectations. However, no sharp enough feature tied uniquely to the existence of the half-flux-quantum state was found. Irregularities in magnetoresistance due to random vortex motion make the observation more difficult. 

%did not help in this regard.  

%\cite{VakaryukVinokur2011}

%%%%%%%%%%%%%%%%%%%%%%%%%%%%%%%%%%%%%%%%%%%%%%

\begin{figure}[t!]
\centering
\includegraphics[natwidth=243,natheight=216]{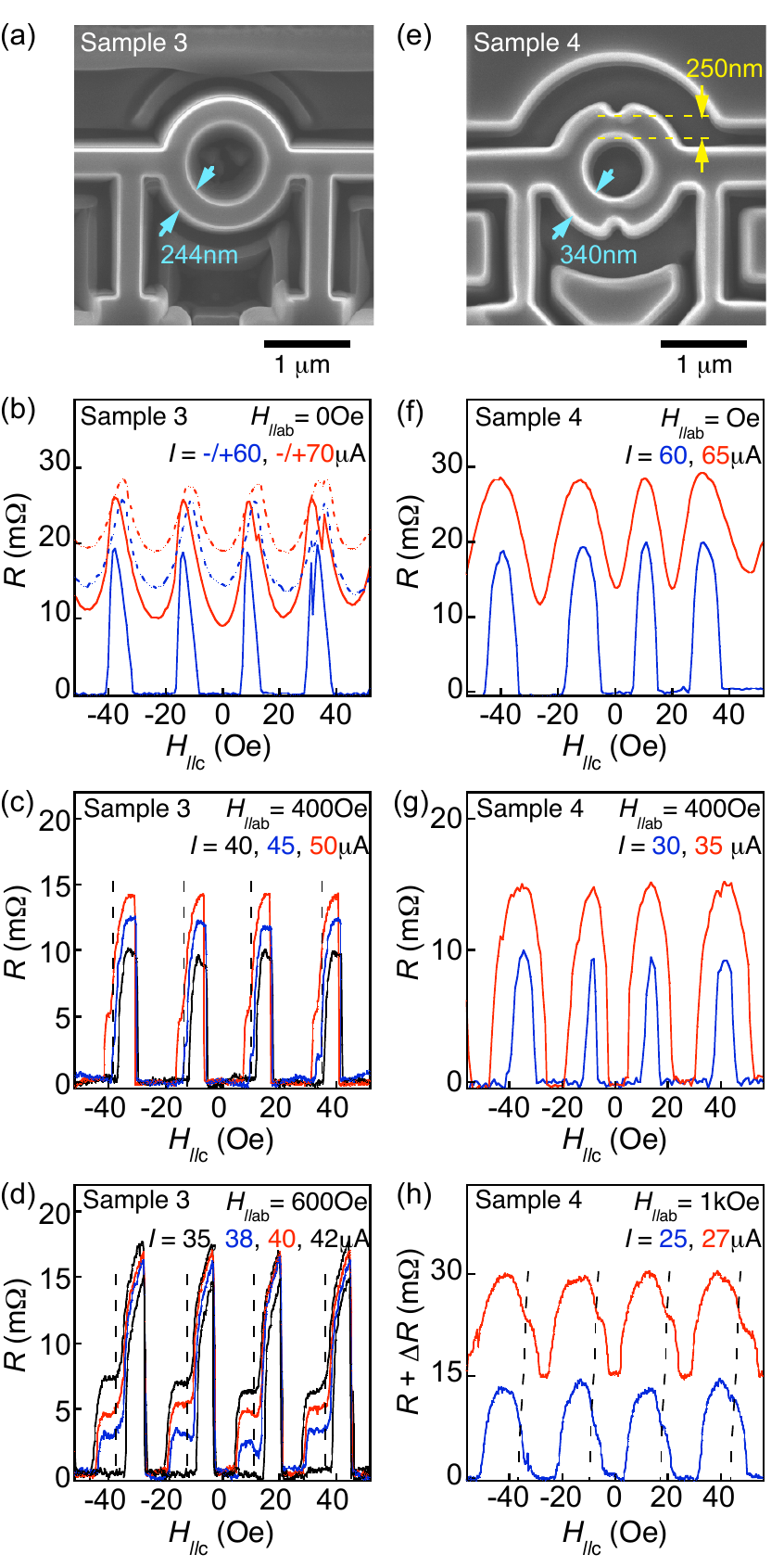}
\caption{
(a) A SEM image for a \Sr\ cylinder, Sample 3. Magnetoresistance oscillation measurement, $R$ vs. $H_{||c}$, with an in-plane field of (b) 0 Oe at $T$= 0.6~K, and (c) 400 Oe and (d) 600 Oe at $T$= 0.3 K for the sample in (a). The measurement currents are indicated starting from the bottom to the top curves. The dash and solid curves in (b) indicate results for positive and negative measurement currents, respectively. 
(d) A SEM image for a cylinder sample with two constrictions. $R$ vs. $H_{||c}$ 
measured at $T$=0.3 K for the sample in (e) with an in-plane field of (f) 0Oe, (g) 400 Oe and (h) 1000 Oe. The top curve shifted vertically by 15 m$\Omega$ for clarity in (h). }
\label{opts}
\end{figure}

The stability condition described above suggests that thinning the cylinder wall, which is limited by FIB damage that could make the sample nonsuperconducting, favors the emergence of the half-flux-quantum state. 
We observed unexpected features in Sample 3 with a mean radius $r\approx$ 545 nm, mean wall thickness $w\approx$ 244 nm and out-of-plane thickness $t\approx$ 482 nm (Fig.~4a). The resistance oscillations are smooth and show little hysteresis even under a substantial in-plane field.
%The sample resistance was found to rise at one side of the peak more sharply than on the other side, with a nearly vertical resistance rise on one side when the measurement current was relatively small (Fig.~4b-4d).  % the higher $H_{||c}$ side of the resistance peaks 
The sample resistance was found to rise gradually only at one side of the resistance peaks, with a nearly vertical resistance rise on the other side when the measurement current was relatively small (Fig.~4b-4d). 
A gradual resistance rise started to appear on the other side when the measurement current was further increased to a critical value, suggesting the critical currents of the two arms of the cylinder are different.
The critical value was found to be around 60~$\mu$A for $H_{||ab}$=400~Oe. 
We estimate that the difference in the critical currents of two arms of the cylinder is on the order of 10~$\mu$A. 
The measured oscillation period is about 23.7 Oe, which corresponds to an effective radius of 527 nm based on $\Delta H =\Phi_0/(\pi r^2)$, smaller than the value obtained from the SEM image. Dip features in the magnetoresistance oscillations were observed at relatively low in-plane fields, 400 Oe and 600 Oe (Figs.~4c and 4d). Increasing the in-plane field appears to promote the two-peak feature, consistent with the trend for the stability region of the \HQV\ state. We speculate that features observed in Sample 3 are the consequence of the \HQV\ state even though trapping of vortices in part of the sample may also generate secondary peaks in the magnetoresistance oscillations. In a study of NbSe$_2$ mesoscopic loops, the field for trapping a vortex in the loop structure was shifted linearly with the measurement current~\cite{MillsNbSe2Loop} while the field position for the observed resistance peaks in Fig.~4d was found to be insensitive to the measurement current.

The uneven critical currents and the smooth magnetoresistance oscillations found in Sample 3 suggest that this sample may possess a weak link 
in one arm of the hollow cylinder due to a sample specific reason.
Theoretical studies suggest that adding physical constrictions in the superconducting hollow cylinder, the kinetic energy of the spin current relative to 
that of the charge supercurrent is determined by the narrowest part of the superconducting loop, $\beta=rw_{min}/2\lambda^2$.~\cite{VakaryukDissertation} 
An additional advantage of the sample with the weak link is that the vortex crossing can be directed to weak link and therefore avoid variation 
in the free energy barrier along the cylinder circumference, thereby increasing the amplitude of the useful magnetoresistance signals.\cite{MillsNbSe2Loop} 
%In the present case, the depinng of parallel vortices is also likely to decrease in samples with a weak link as the needed measurement current also tends to be small. All these factors suggest that a sample featuring one or two weak links may favor the observation of the half-flux-quantum states.

In Fig.~4e-4h, results from a sample featuring two constrictions were shown. The sample has a mean radius $r\approx510$ nm, a wall thickness of 340 nm, 
and the width of the constriction roughly 250 nm. Since the width of the constrictions in this sample is comparable to the wall thickness of Samples 1 and 2, 
based on the free-energy consideration, the \HQV\ state should appear at a similar in-plane field. Consistently, no feature was seen in the magnetoresistance oscillations at an in-plane field of 400 Oe (Fig.~4g). 
%Even when the in-plane field was increased to as large as 1000 Oe, the magnetoresistance peaks remain smooth. 
After an in-plane field of 1000 Oe was applied was a dip found in resistance peaks (Fig.~4h), with features similar to that seen in Sample 3,
consistent with expectations from the above analysis. 

%It is interesting to note that the stability region for the half-flux-quantum state seen in the magnetometry experiment was usually not centered at half applied flux quanta,~\cite{Jang2011Science} consistent with features seen in our magnetoresistance oscillations.

%%%%%%%%%%%%%%%%%%%%%%%%%%%%%%%%%%%%%%%%%%%%

To conclude, results from the magnetoresistance oscillation measurements suggest that, even though the \HQV\ state may indeed exist 
in micron-sized, doubly connected cylinders of single crystal \Sr, the free energy barely vary as a function of applied flux enclosed in the cylinder within 
the stability regime. In the torque magnetometry measurements, a plateau in the magnetization can be detected as long as the free energy 
of the half-flux-quantum state is lower than that of the full-flux-quantum state. The difference of the free energy between them is difficult 
to determine through the analysis of the measured magnetoresistance oscillations because the connection between 
the rate of the vortex crossing and the free energy barrier is complicated. Our measurements show that the kinetic energy part of the free energy, 
which is tuned by the applied magnetic flux, is only a small part of the total free energy. The spin-orbital coupling 
to the free energy, which should be large in \Sr,\cite{Kee2000PRB} may have made the kinetic part of the free energy even smaller, and the detection 
of half-flux-quantum through magnetoresistance oscillation measurements significantly more difficult than the torque magnetometry measurements.

We would like to thank H-Y. Kee, J. Kirtley, Y. Maeno, Z. Wang, Y. Xin, Z. Xu, C-C. Tsuei, S-K. Chung, J. K. Jain, C. Kallin, A. J. Leggett, 
J. A. Sauls, M. Sigrist, V. Varkaruk, K. Roberts, S-K. Yip, B. Zakrzewski and S. Mills for useful discussions. The work done at Penn State is supported by DOE 
under Grant No. DE-FG02-04ER46159, at SJTU was supported by MOST of China (2012CB927403). at Tulane by the U.S. Department 
of Energy under EPSCoR Grant No. DE-SC0012432 with additional support from the Louisiana Board of Regents (support for materials 
synthesis and characterization).

%%%%%%%%%%%%%%%%%%%%%%%%%%%%%%%%%%%%%%%%%%%%%%%%%%%%%%%%%%%%%%%%%%%%%%%%

%\bibliography{XCaiBib}

\end{document}